# $^{111}$Cd- TDPAC study of pressure-induced valence transition in $YbAl_2$

**A.V. Tsvyashchenko, L.N. Fomicheva**

*Vereshchagin Institute for High Pressure Physics, RAS, 142190 Troitsk, Russia*

**V.B. Brudanin, O.I. Kochetov, A.V. Salamatin, A. Velichkov**

*Joint Institute for Nuclear Research, P.Box 79, Moscow, Russia*

**M. Wiertel, M. Budzynski**

*Institute of Physics, M. Curie-Sklodowska University, 20-031 Lublin, Poland*

**A.A. Sorokin, G.K. Ryasniy, B.A. Komissarova**

*Skobeltsyn Institute of Nuclear Physics, 119899 Moscow, Russia*

**M.Milanov**

*Institute for Nuclear Research and Nuclear energy, BAS, 1000 Sofia,Bulgaria*

The quadrupole interaction at $^{111}$Cd impurity nuclei in the intermediate-valence compound $YbAl_2$ has been measured under pressure up to 80 kbar by the TDPAC spectroscopy. It was found that the quadrupole frequency $\nu_Q$ measured on the $^{111}$Cd located at the Al sites in $YbAl_2$, varies nonlinearly and increases by almost four times with the pressure increase up to 80 kbar. A linear correlation between the mean Yb valence, derived from Yb $L_3$ OFY-XAS and RXES measurements, and the electric field gradient (the quadrupole frequency $\nu_Q=eQV_{zz}/h$) has been observed. The quadrupole frequencies on $^{111}$Cd in the $GdAl_2$, $YbAl_3$, $TmAl_3$ and $CaAl_2$ compounds have been measured, also. The possibility of determining the valence of Yb in the Yb compounds with *p*-metals from the relation $\nu_Q(\upsilon(P)) = \nu_2 + (\nu_3 - \nu_2)\,\upsilon(P)$ has been considered.

## INTRODUCTION

Ytterbium intermetallic compounds exhibit interesting physical phenomena such as intermediate valence, Kondo, or heavy fermion behavior[1]. These properties are highly sensitive to chemical environment as well as to external pressure or temperature. These is due to the fact that in the atomic ground state Yb is divalent with a filled $f^{14}$ shell, but in the solid state the *f*-electrons may play an active role in the formation of the electronic band structure. The accurate theoretical description of the Yb compounds remains a challenge. Nevertheless, by applying self-interaction corrections (SIC)[2] in the local-spin-density approximation (LSDA) , the authors could give a valid description of the effective valences of Yb in the Yb compounds. In this description, for a nominal trivalent case, where Yb possesses a localized $f^{13}$ shell, the 14$^{th}$ *f*-electron forms a very narrow band situated among normal *s*-*d*-derived conduction states. The Fermi level is pinned to this narrow band that becomes partially occupied. As a consequence, the total *f* occupancy is non-integer and falls in between 13 and 14, and the nominally trivalent Yb configuration therefore describes an intermediate valent state with the number of the band $f$ - electrons, $n_f$ $(0 \leq n_f \leq 1)$.

The conventional definition of the effective valence for rare earths identifies the valence by the number of the band $f$ -electrons having passed into the $5d$ -electron band (hereinafter they are referred to as "band $f$ -electrons" in inverted commas), $1 - n_f = n_{5d}$, or the number of $f$ holes $1 - n_f = n_h$ [1]. Thus, the Yb valence is determined as $\upsilon = 3 - n_f = 2 + n_{5d} = 2 + n_h$.

In order to obtain quantitative experimental information on the mean valence, researchers widely employ the following techniques: $L_{III}$ – edge x-ray-absorption spectroscopy (XANES)[3], Mössbauer effect (ME) (the mean valence derived from isomer shift measurements)[4] and photoemission spectroscopy of the $4f$- levels[5]. It should be noted that the parameters, measured by these techniques, couldn't be calculated from the first principles.

On the other hand, the electric field gradient (EFG) is a ground-state property of a solid and depends sensitively on the asymmetry of the electronic charge in a crystal. Blaha *et al.*[6] developed a first-principles method to compute the EFGs. They have used the full-potential linear augmented-plane-wave (FL-LAPW) method and have calculated the EFG directly from the self-consistent charge density by solving the Poisson's equation.

But since the formation of the electron bands by hybridization between the states of different shells leads to a non-integer occupation of the related shells and to the redistribution of the charge density, one can expect that the EFG may also reflect the change of the valence of rare earth ions.

Indeed, the experimental studies of the electric (nuclear) quadrupole interaction (EQI) preformed by the time-differential perturbed angular correlation (TDPAC) method in a series of intermetallic compounds $RIn_3$[7], $RSn_3$[7] and $RAl_2$[8] (R = a rare earth element) demonstrated a sharp drop of the EFG at $^{111}$Cd in $YbIn_3$, in $RSn_3$ with R = Eu and Yb, and in $EuAl_2$ with regard to other trivalent rare earths $R^{+3}$. This drop was assigned to be due to the difference in the valence between Eu and Yb with $\upsilon = 2$ and the compounds with $R = R^{+3}$.

When performing *ab initio* numerical analysis[9] of the experimental data[7] on the EFG of $^{111}$Cd in $YbIn_3$, $YbSn_3$ and $EuSn_3$, the authors used the FL-LAPW method [6,10]. They showed that the increased $V_{zz}$ value in $R^{+3}In(Sn)_3$ relative to the EFG values observed in $R^{+2}In(Sn)_3$ is caused by an additional polarization of Cd $p$-electrons due to a strong $f$ - $p$ hybridization near the $xy$ plane[9].

For all $R^{3+}Al_2$ compounds it was shown[8] that the EQI of $^{111}$Cd is characterized by the quadrupole coupling constant (the quadrupole frequency) $\nu_Q \approx 50$ MHz, slightly dependent on $Z$ of a rare earth element, whereas in $Eu^{+2}Al_2$ it is several times lower, $\nu_Q \cong 8$ MHz. No data for $YbAl_2$ were reported in [ref. 8]. However, the $^{151}$Eu Mössbauer studies[11] under hydrostatic pressure up to 410 kbar revealed that – contrary to expectations – the $4f$ electrons in $EuAl_2$ remain fully localized, that is, the Eu ions are divalent, even at the reduction of the unit cell volume by ~ 30%.

On the other hand, the resent results[12] of the pressure-induced valence transition investigations performed by Yb $L_3$ OFY-XAS and RXES for $YbAl_2$ have revealed that the Yb valence increases from 2.25 at ambient pressure to 2.8 at 80 kbar. In so doing, preliminary high-pressure x-ray diffraction studies[13] had found no changes in the crystal structure of $YbAl_2$ up to 80 kbar.

In view of the said above, in this work for studies by the $^{111}$Cd -TDPAC method of a pressure-induced valence change we chose the $YbAl_2$ compound crystallizing in the cubic structure of a $MgCu_2$ type. The experimental determination of the pressure dependence of the quadrupole coupling constant $\nu_Q=eQV_{zz}/h$ (the quadrupole frequency) measured on $^{111}$Cd probes by the time-differential perturbed angular correlation method enabled us to determine the linear dependence of the EFG on the number of the "band-$f$-electrons", or $1 - n_f = n_h = n_{5d}$. We demonstrated that this dependence can be used for determining the valence of Yb ions for the $YbAl_3$ and $YbIn_3$ compounds.

## EXPERIMENT

The measurements were carried out by the time-differential perturbed angular correlation method using the 171-245 keV $\gamma$-ray cascade in $^{111}$Cd populated through the 2.8 d isotope $^{111}$In electron capture decay. The cascade proceeds via the 245 keV level with the half-life $T_{1/2}$ = 84 ns, spin $I$ = 5/2, and quadrupole moment $Q$ = 0.83 b.

The $^{111}$In activity was produced via the $^{109}$Ag ($\alpha$, 2$n$) $^{111}$In reaction through irradiating a silver foil with the 32 MeV $\alpha$-beam. The nuclear probe $^{111}$In-$^{111}$Cd was introduced into the lattice of $YbAl_2$, $YbAl_3$, $GdAl_2$, $TmAl_3$, and $CaAl_2$ by the high pressure synthesis: the constituents (Yb, Tm, Gd, Ca and Al) taken in proper amounts with an overall weight of about 500 mg were melted together with a small piece of the irradiated silver foil (≈0.5 mg) in a special chamber under pressure of 80 kbar [14].

The X-ray diffraction of the powdered samples evidenced that the alloys were single phase; $YbAl_2$ had the cubic Laves phase structure, and $YbAl_3$, $TmAl_3$ had the cubic $Cu_3Au$ structure. Our previous experience has proved that this procedure ensures the substitution of $^{111}$In for Ru in the Laves compounds $LaRu_2$ and $CeRu_2$ [14], and, as will be shown below, for Al in $GdAl_2$[8], and, presumably, in $YbAl_2$, $YbAl_3$, $CaAl_2$ and $TmAl_3$.

The TDPAC measurements were carried out using a 4-detector spectrometer equipped with a small-size hydraulic four-arm press of the capacity up to 300 $t$ [15].

As the samples were polycrystalline and paramagnetic at room temperature, the perturbation of the angular correlation can be described by the perturbation factor for the static electric quadrupole interaction [16]:

$$G_{22}(t; \nu_q, \eta, \delta) = s_{20} + \Sigma s_{2n} cos(\omega_n t) exp(-1/2 \delta \omega_n t).$$

The hyperfine frequencies $\omega_n$ depend on the quadrupole coupling constant $\nu_Q=eQV_{zz}/h$ (the quadrupole frequency), and the asymmetry perameter $\eta=(V_{xx}-V_{yy})/V_{zz}$, where $V_{ii} = \partial^2 V/\partial i^2$ ($i = x,y,z$) are the principal-axis components of the EFG tensor. The coefficients $s_{2n}$ depend only on $\eta$ ($1 \geq \eta \geq 0$). For the nuclear spin $I$=5/2, $n$=1, 2, 3. The exponential factor accounts for possible random lattice defects, and δ is the relative half-width of the Lorentzian distribution. Here we restrict ourselves to the perturbation parameter of the second order since the unperturbed angular correlation coefficient $A_{44} << A_{22}$ ($A_{22}$ = - 0.18).

The perturbation factor $G_{22}(t)$, describing a nuclear spin precession due to a hyperfine interaction, was determined in a usual way from the angular anisotropy, obtained by combining the delayed coincidence spectra measured at the angles of 90° and 180° between detectors, $N(90^o,t)$ and $N(180^o,t)$, through the expression $R(t) = -A_{22}Q_2G_{22}(t)$, where

$R(t) = 2[ N(180^o,t) - N(90^o,t)] / [ N(180^o,t) + N(90^o,t)]$, $Q_2 \approx 0.80$ is the solid-angle correction.

A 100 mg sample of $YbAl_2$ doped with $^{111}$In was positioned inside a rock-salt ampoule that was used as a pressure-transmitting medium. The high pressure was generated in a calibrated "toroid"-type device[17]. The calibration of the device was confirmed through measuring the EQI of $^{111}$Cd in $^{111}$In-doped metallic Zn at normal pressure and at nominal pressure of 30 *kbar*. The obtained values, $\nu_Q$=131 and 113 MHz, respectively, were in good agreement with those reported by da Jornada and Zavislak [18].

## RESULTS AND DISCUSSION

The $^{111}$Cd - TDPAC spectra $R(t) = -A_{22}Q_2G_{22}(t)$ observed in the $YbAl_2$ and $GdAl_2$ compounds at different pressures are presented in Fig.1 and 2.

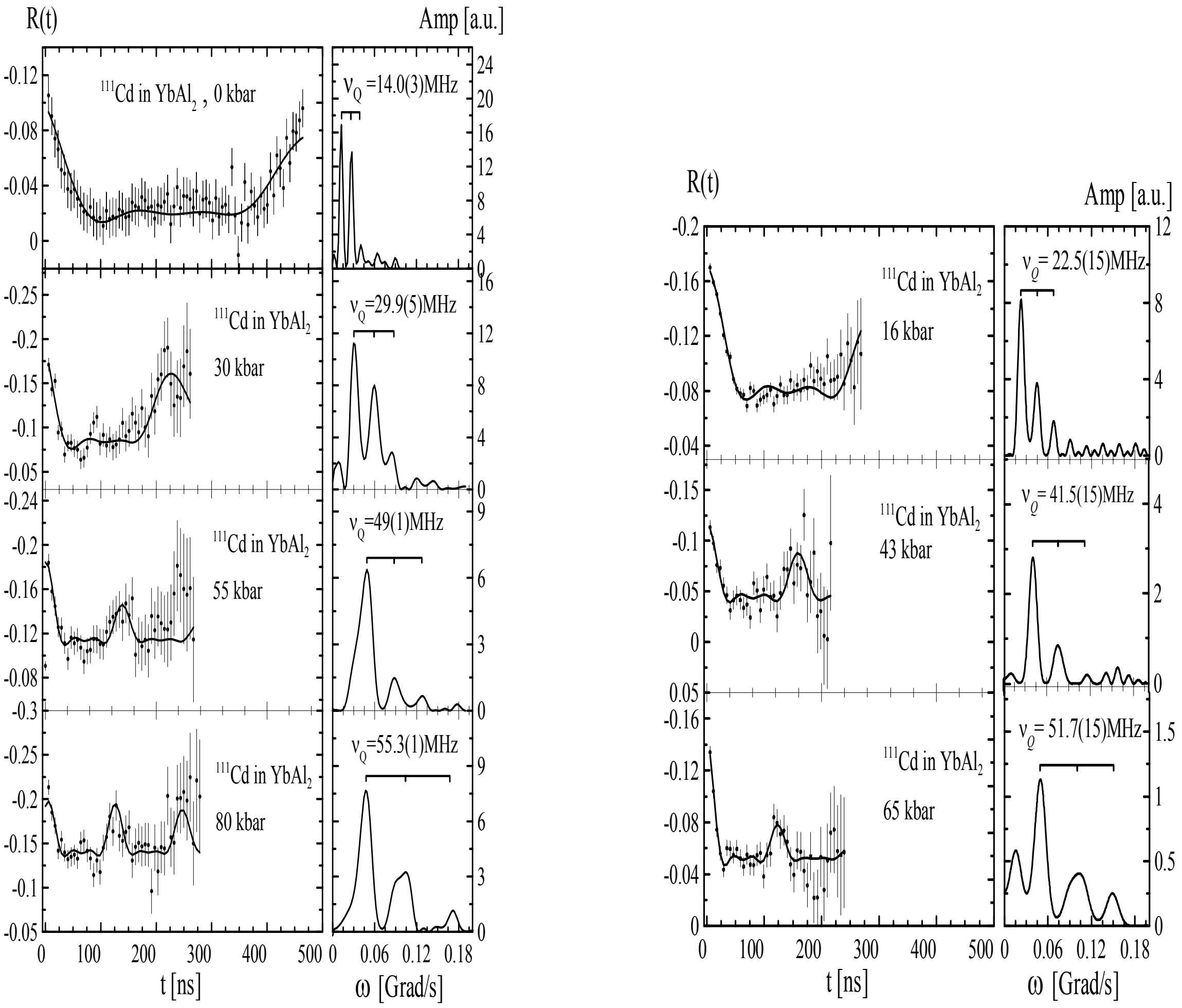

Fig. 1. Time spectra of the angular correlation anisotropy, R(t) (left panel) and their Fourier transform (right panel), for $^{111}$Cd in $YbAl_2$ measured at various pressures. The increase in the deviation of the anisotropy spectrum with the increase of pressure from a zero line is due to the scattering in the sample that varies in shape from a cylinder to a thin disk (see ref. 19) as pressure rises.

The TDPAC spectrum for $YbAl_2$ measured at normal pressure can be best fitted with $\nu_Q$=14.0(3) MHz assuming the asymmetry perameter $\eta = 0$. Here we note in advance that in all cases the zero pressure $R(t)$ curves were the same before and after compression, and the values of the quadrupole frequency (QF) $\nu_Q$ extracted from the experimental $R(t)$ curves coincided. The EQI parameters measured at normal pressure for $GdAl_2$, $\nu_Q$ = 51(1) MHz, $\eta = 0$, were in good agreement with the data reported in [ref.8]. As was mentioned, this result confirms that in the sample prepared by the synthesis at high pressure through the method described above, $^{111}$In substitutes for Al.

The value of the quadrupole frequency $\nu_Q$ for $YbAl_2$ is several times lower than that for $GdAl_2$ (where the Gd ion is trivalent); the same was earlier observed for $EuAl_2$[8].

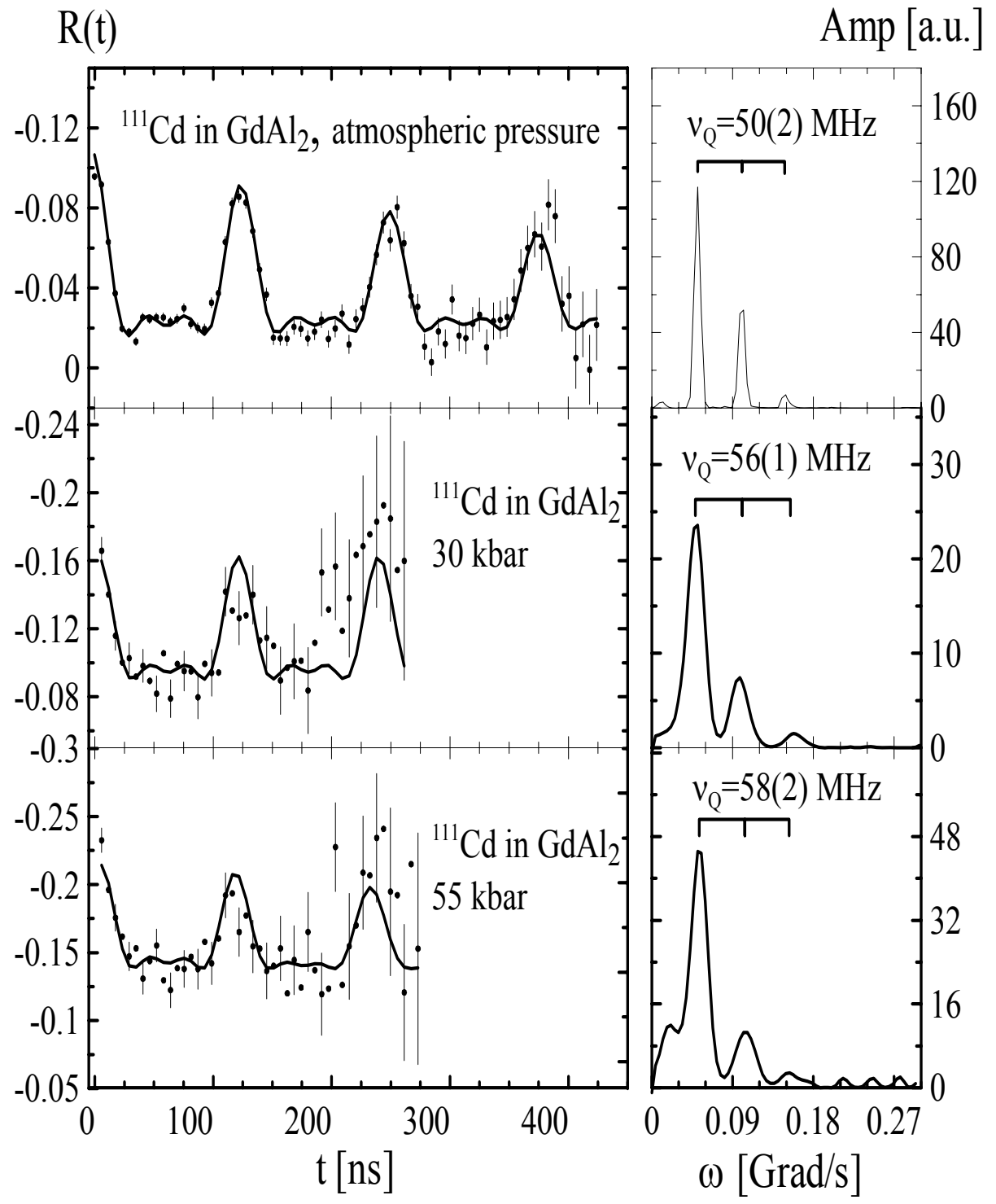

Fig. 2. Time spectra of the angular correlation anisotropy, R(t) (left panel) and their Fourier transform (right panel), for $^{111}$Cd in $GdAl_2$ measured at various pressures.

The variations of the QF $\nu_Q$ with pressure for $^{111}$Cd in $YbAl_2$ and $GdAl_2$ are shown graphically in Fig. 3.

As seen from the pressure dependence of the quadrupole frequency $\nu_Q$ measured on the $^{111}$Cd impurity nuclei located at the Al sites in $YbAl_2$ (see Fig.3), the QF varies nonlinearly and increases by almost four times with the pressure increase up to 80 kbar. For $GdAl_2$ the QF rises with the same pressure increase for not more that 10%. A similar behavior was observed for the pressure-induced change in the isomer shift of $EuNi_2Ge_2$ pointing to a valence transition from $Eu^{+2}$ to $Eu^{+3}$ around 50 kbar in [ref. 20].

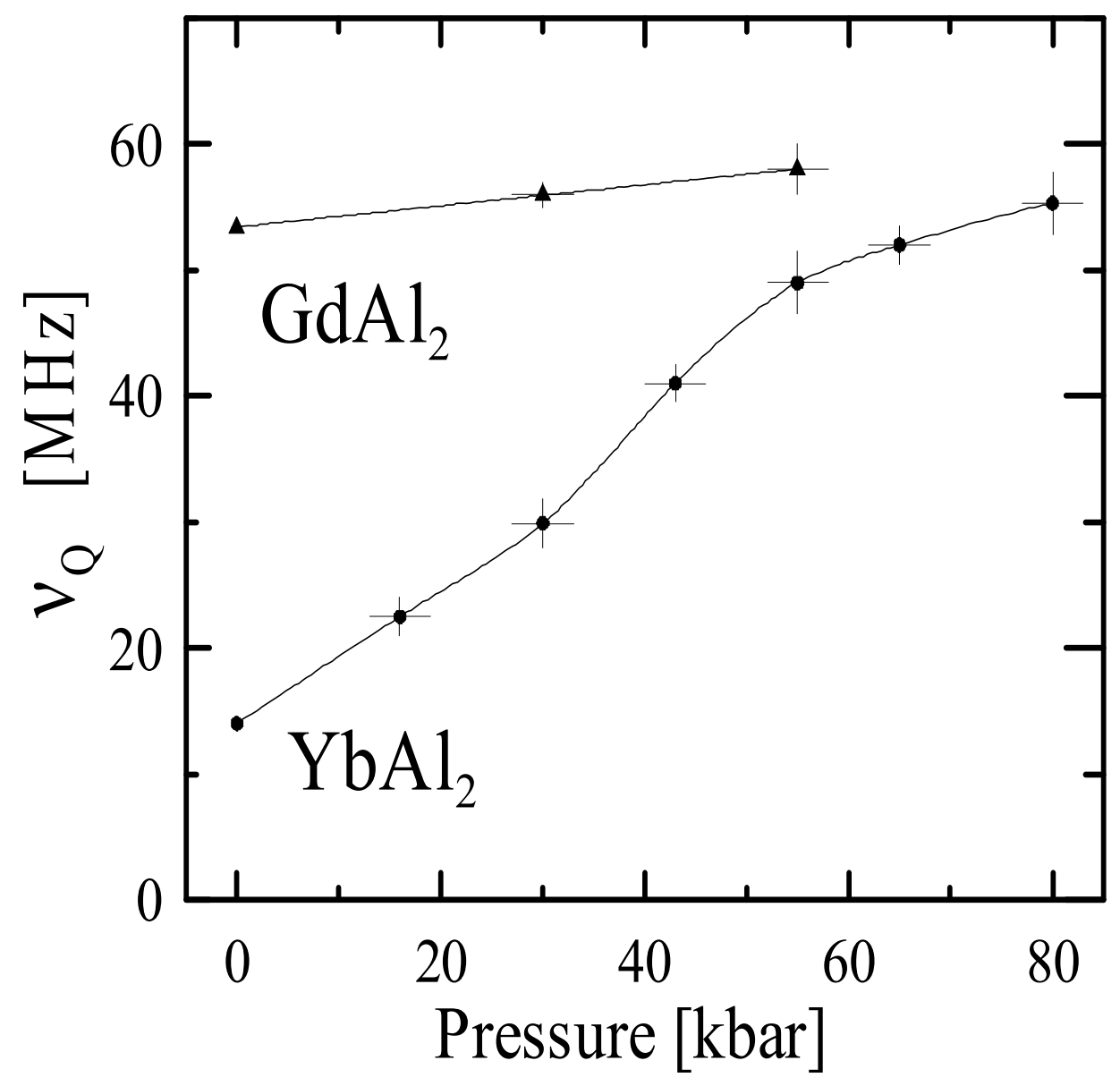

Fig. 3. Pressure dependence of the quadrupole frequency $\nu_Q$ for $^{111}Cd$ in $YbAl_2$ and $GdAl_2$.

This considerable difference in the QF values and behaviour according to pressure for these particular compounds can be explained by the folowing fact. The ytterbium ion valence in $YbAl_2$ at normal pressure is 2.25 and it rises to 2.8 [12] with the pressure increase to 80 kbar. In $GdAl_2$, the gadolinium ion valence is 3 and as the pressure in this range goes higher, it does not change. Since the valency change ($\upsilon = 2 + n_d = 2+(1- n_f)$) is determined by the filling of the 5*d*- band by the transition of *f*-electrons, both the QF value and QF variation according to pressure is governed by the fact that the main contribution to the electric field gradient is made by the "band-*f*–electrons". If the number of the "band-*f*-electrons" remains the same, then the increase of the EFG can occur due to the reduction of the lattice parameter, as was shown in [Ref. 7] for cubic compounds $RIn_3$ . This can explain an inessential pressure-dependent rise of the QF, observed for GdA12.

Using the data obtained in [Ref. 12] for the pressure dependence of the ytterbium ion valence, we have plotted the dependence of the EFG at $^{111}Cd$ in $YbAl_2$ on the number of the "band *f*-electrons" ($1 - n_f = n_{5d}$) and have found this dependence to be linear (see Fig.4).

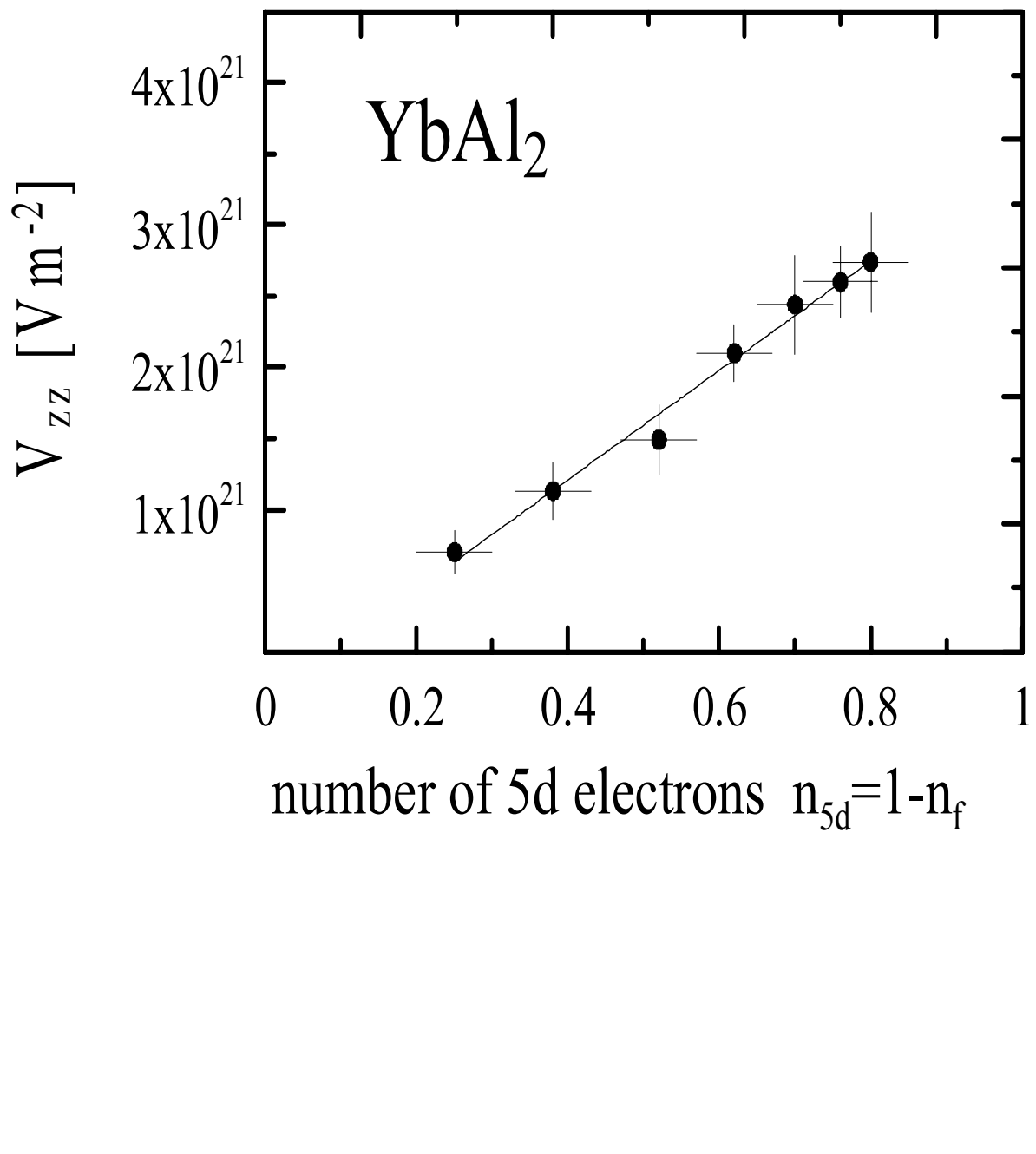

Fig. 4. The electric field gradient ($V_{zz} = v_Q h / eQ$) for $^{111}$Cd in $YbAl_2$ versus the number of the band $f$-electrons having passed into the 5d-band ($1 - n_f = n_{5d}$).

Thus, based upon a linear dependence of the EFG (or the quadrupole frequency which is directly measured in the experiment) on the number of the "band $f$-electrons" ($1-n_f$), one can determine the valence [$\upsilon = 2+(1-n_f) = 2 + \upsilon(P)$] of the ytterbium ions by using the following relation: $\nu(\upsilon(P)) = \nu_2 + (\nu_3 - \nu_2)\,\upsilon(P)$, where $\nu(\upsilon(P))$ - is the QF measured on $^{111}$Cd in a ytterbium compound with an $p$-metals (for example Al, In); $\nu_3$ – is the QF measured in a similar compound with a trivalent ytterbium or a trivalent rare earth; $\nu_2$– is the QF measured in a similar compound with divalent ytterbium or with another divalent metal, for example, Eu, or Ca. This relation is similar to the one determining the valence for Eu from isomeric shift [4,21].

Indeed, for $YbAl_2$ from a linear dependence of the QF, $\nu_2$ is equal to –5,8(4) MHz and the value of ($\nu_3 - \nu_2$) is equal to 76,3(6) MHz. As ytterbium in the $YbAl_2$ compound at normal pressure has a non-integer valence, for the evaluation of $\nu_2$ we have measured the QF in $CaAl_2$. This compound has the same crystal structure as $YbAl_2$ , but Ca is usually regarded as a divalent analogue of Yb. From the $^{111}$Cd -TDPAC measurements we found the value $\nu_2$ for $CaAl_2$ to be equal to 5,7(2) MHz (see fig.5.)and close to the $\nu_2$ value defined from QF linear dependence for $YbAl_2$.

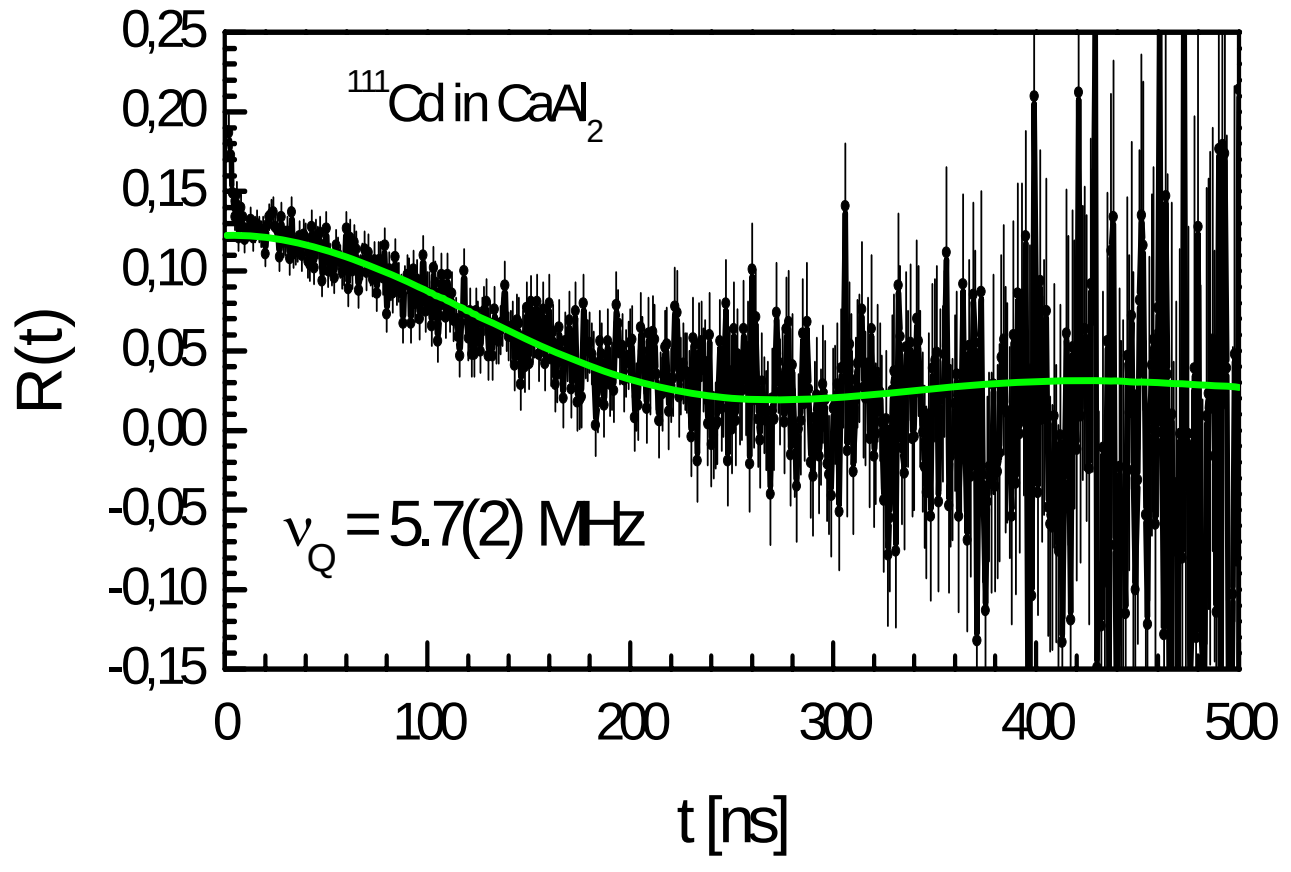

Fig.5. Time spectrum of the angular correlation anisotropy, R(t) for $^{111}$Cd in $CaAl_2$ measured at normal pressures.

It is also known[22] that by employing the X-ray $L_{111}$- absorption technique, the Yb ion valence for the $YbAl_3$ compound was determined and found equal to 2,83 at room temperature (i.e., $\upsilon(P) = (1-n_f) = 0{,}83$). To verify our assertion that the ($1-n_f$) - quadrupole frequency dependence is linear for other compounds, we have performed the measurements of the QF for the $YbAl_3$ and $TmAl_3$ compounds that have a $Cu_3Au$ – type crystalline structure. Angular anisotropy spectra for these compounds are given in Fig.6.

It turned out that for $YbAl_3$, the QF value is equal to 73.3 MHz, and for $TmAl_3$ it is 85,2 MHz. To determine the valence of the Yb ions in this compound we employ the relation suggested above. We assume that the value ( $\nu_3$ - $\nu_2$) does not depend on the crystalline structure and remain equal to 76,3 MHz, as it was found for $YbAl_2$ , while the $\nu_2$ value do depend on the crystalline structure of the compound.. Assuming that $\nu_3$ $(Yb^{+3}Al_3)$ = $\nu(Tm^{+3}Al_3)$= 85,2 MHz, we can easily determine $\nu_2(Yb^{+2}Al_3)$ = 8,9 MHz. Then for $YbAl_3$ the value $\upsilon(P) = [\nu(YbAl_3) - \nu_2(Yb^{+2}Al_3)]/ (\nu_3 - \nu_2) = (73.3\text{-}8.9)/ 76.3 = 0.84$.
The ratio of these frequencies gives us the value $\upsilon(P)$ = 0,84 and, accordingly, the valence 2.84 for Yb ions in $YbAl_3$. Thus, the valence determined through the TDPAC method is in good agreement with the data[22] on the valence of Yb in $YbAl_3$. If we use this line of reasoning for $YbIn_3$, then, from the data[7] on the quadrupole frequency measurements by $^{111}$Cd-TDPAC (for $YbIn_3$ $\nu_Q$ = 38,7 MHz, and for $TmIn_3$ $\nu_Q$ = 87,7 MHz), we obtain the valence of a Yb ion in $YbIn_3$ equal to 2,36 and the value $\nu_2(Yb^{+2}In_3)$ = 11.4 MHz. This is close but not similar to the value of a Yb ion ($\upsilon$ =2,50) in [Ref. 23]. The difference can be accounted for by the fact that the valence in [Ref. 23] was found without using the $L_{III}$ – edge x-ray-absorption spectroscopy (XANES) technique.

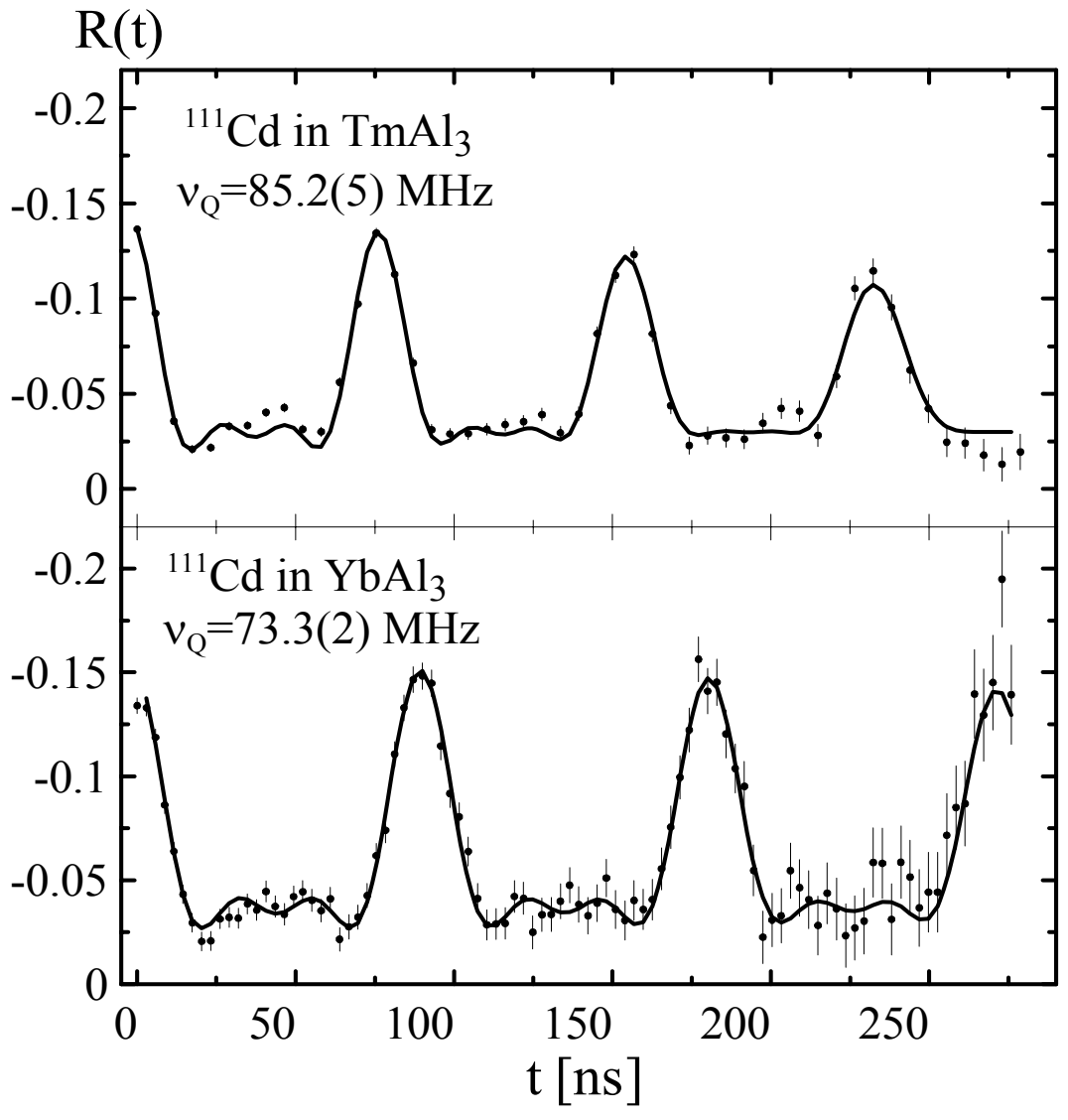

Fig. 6. Time spectrum of the angular correlation anisotropy, R(t) for $^{111}$Cd in $TmAl_3$ and $YbAl_3$, measured at normal pressures.

## CONCLUSION

In this way, the results obtained in this work make it possible to state that the variation of the EFG (the quadrupole frequency $\nu_Q = eQV_{zz}/h$) for the Yb compounds with $p$ –metals is linearly connected with the number of the $f$ –electrons having changed over to a conduction band, i.e., with the number (1-$n_f$). This regularity allows the determination of the Yb valence through the $^{111}$Cd-TDPAC method.

In the work by Svane et. al [2] when calculating the Yb ion valence from the first principles, a linear correlation was also determined between the difference of the energies of the valent states and the number (1-$n_f$) of the electrons having come over to a conduction band. However, since the change of the energies between two different valent states is caused by the redistribution of *f*-electrons, which, in its turn, causes the variation of the EFG, one can speak of an agreement between the theory[2] and experimental results obtained in the present work. Although for drawing a correct comparison, theoretical calculations of the EFG vs Yb valence dependence are necessary.

## ACKNOLEGEMENTS

This work was supported by the grants of the Russian Foundation for Basic Research (project nos. 07-02-00280) and by special programs of the Department of Physical Science, Russian Academy of Sciences. The work at the Joint Institute for Nuclear Research was carried out under the auspices of a Polish representative in the JINR.